\documentclass[manuscript]{emulateapj}  

\begin{document}


\title{Did Massive Primordial Stars Preenrich the Lyman Alpha Forest?}

\author{Michael L. Norman\altaffilmark{1}, Brian W. O'Shea\altaffilmark{1,2} 
\& Pascal Paschos \altaffilmark{1,3}}

\altaffiltext{1}{Center for Astrophysics and Space Sciences,\\  
University of California at San Diego, La Jolla, CA 92093, U.S.A.\\ 
Email:  mnorman, bwoshea, ppaschos@cosmos.ucsd.edu}

\altaffiltext{2}{Department of Physics,\\
University of Illinois in Urbana-Champaign}

\altaffiltext{3}{Department of Astronomy,\\
University of Illinois in Urbana-Champaign}


\begin{abstract}

We examine the dynamical evolution and statistical properties of the supernova ejecta of 
massive primordial stars in a cosmological framework
to determine whether this first population of stars could have enriched the
universe to the levels and dispersions seen by the most recent observations of
the Lyman-$\alpha$ forest.  We evolve a $\Lambda$CDM model in a 1 Mpc$^3$ volume
 to a redshift of $z = 15$ and add 
``bubbles'' of metal corresponding to the supernova ejecta of the first generation of 
massive stars in all dark matter halos with masses greater than $5 \times 10^5
M_{\odot}$.  These initial conditions are then evolved to $z = 3$ and the 
 distribution and levels of metals are compared to observations.  
In the absence of further star formation the primordial metal is 
initially contained in halos and filaments.  Photoevaporation of metal-enriched 
gas due to the metagalactic ultraviolet background radiation at the epoch of 
reionization ($z \sim 6$) causes a sharp increase of the metal volume filling 
factor.  At $z = 3$, $\sim 2.5\%$ of the simulation volume ($\approx 20\%$ of 
the total gas mass) is filled with gas enriched above a metallicity of 
$10^{-4} Z_{\odot}$, and less than $0.6\%$ of the volume is enriched
above a metallicity of $10^{-3} Z_{\odot}$.  This suggests that, even with
the most optimistic prescription for placement of primordial supernova and
the amount of metals produced by each supernova, this population of stars cannot 
entirely be
responsible for the enrichment of the Lyman-$\alpha$ forest to the levels and
dispersions seen by current observations unless we have severely underestimated
the duration of the Pop III epoch.  However, comparison to observations
using carbon as a tracer of metals
shows that Pop III supernovae can be significant contributors to the very low
overdensity Lyman-$\alpha$ forest.
\end{abstract}

\keywords{cosmology: theory -- intergalactic medium -- galaxies: formation}


\section{Introduction}\label{intro}

Recent observations by \citet{Schaye03} have shown that the Lyman-$\alpha$
forest is polluted with metals at very low densities.  The distribution of
metal is very strongly dependent on overdensity, with median metallicity
values ranging from $[C/H] = -4.0$ at log~$\delta = -0.5$ (where $\delta$ is
defined as ($\delta \equiv \rho/\bar{\rho}$) to $[C/H] = -2.5$ at 
log~$\delta = 2.0$ using their fiducial UV background model.  Their
observations show little evidence for metallicity evolution of the Lyman-$\alpha$
forest over the redshift range $z = 1.5-4.5$.

The lack of observed evolution in metallicity is suggestive of a very early 
epoch of stellar evolution.  Recent observations by the Wilkinson Microwave
Anisotropy Probe suggest an epoch of star formation in the redshift range of
$z = 11-30$ \citep{Kogut03}, which is consistent with the simulation 
results of \citet{Abel02} and \citet{bromm02}, which suggest that the 
first generation of stars (known as Population III, or Pop III) formed in 
the redshift range $z = 20-30$.   The Abel et al. results, which are the 
highest-resolution simulations of formation of the first generation of 
primordial stars to date, also suggest 
that Pop III stars are very massive - on the order of $\sim 200 M_{\odot}$.
Stars that are in this mass range will die in extremely energetic 
pair-instability supernovae and can eject up to 57 M$_{\odot}$ of $^{56}$Ni.
  \citep{Heger02, Heger03a}.  The formation site of Pop III stars is in halos with
total masses of $\sim 10^6 M_{\odot}$.  \citep{Abel02, Yoshida03}.
 These halos have escape velocities 
which are on the order of 1 km/s.  Due to the shallowness of the potential wells 
that Pop III stars form in, \citet{Ferrara98} suggests that ejecta from a 
massive Pop III supernova can propagate to very
large distances (far greater than the halo virial radius), a result which is supported
in simulations performed by \citet{Bromm03}.

In this paper we describe the results of cosmological hydrodynamic simulations which
address  whether or not a population of massive primordial
stars can be responsible for metal enrichment of the Lyman-$\alpha$ forest to the
level and dispersion seen today.  We examine the most optimistic possible scenario
for Pop III star formation and enrichment in order to establish an upper limit
on metal enrichment of the Lyman-$\alpha$ forest due to Population III stars.

\section{Simulations}\label{sims}

The simulations were set up using the concordance cosmological model
($\Omega_{m} = 0.3$,$\Omega_{\Lambda}=0.7$, $\Omega_{b}=0.04$ and a Hubble
parameter of $h=0.7$ where h is in units of 100 km s$^{-1}$ Mpc$^{-1}$.  
Initial perturbations in the dark matter and baryon density were created with an
\citet{EisHu99} power spectrum with $\sigma_8 = 0.9$ and $n=1$ using the
Zel'dovich approximation \citep{Bertschinger91} in a box which is 0.7 $h^{-1}$ 
comoving Mpc on a side. In our simulations we use a computational box with 
$256^3$ grid cells with a comoving spatial resolution of $2.7 h^{-1}$ kpc
 and a dark matter mass resolution of 1477 $h^{-1}$ $M_{\odot}$.

The simulation was initialized at z=99 and was allowed to evolve to z=15 
using the Eulerian adaptive mesh refinement code Enzo \citep{Bryan99}.
The simulation was stopped at $z=15$ and all halos with dark matter mass
M$_{DM} \geq 5 \times 10^{5}$ M$_{\odot}$ were found using the HOP halo-finding
algorithm \citep{EisHut98}.  This yielded 184 halos, each of which we assume
produces one pair instability supernova.  Note that we are ignoring negative
feedback, which raises the minimum halo mass which can form Pop III stars
\citep{Machacek01}.  This is consistent with our desire to simulate a best-case
scenario.  We discuss this and other assumptions in Section~\ref{discussion}.

At this point, two separate cases are
considered.  In the first case (referred to as Case `A'),
 spheres of uniform metal density 1 kpc (proper)
in radius with 127 $M_{\odot}$ of metal are placed in the simulation, centered on
the cell of highest dark matter density of each halo. This corresponds to a mass 
averaged metallicity
in the volume of $<Z> \equiv M_{Z}/M_{B} = 4.02 \times 10^{-5}$ $Z_{\odot}$ 
(where $M_Z$ and $M_B$ are total metal and baryon masses in the simulation volume), 
which remains constant
throughout the simulation. No other modifications to the data were made -- in particular,
the baryon density, temperature and peculiar velocities and dark matter
density were unmodified.

In the second case (Case `B'), the spheres of uniform metal density are placed down in the
same manner. In addition, the baryon gas in the corresponding volume is smoothed
to the cosmic mean ($<\rho> = \Omega_b \rho_{c}$), and the temperature of the
baryon gas is set to 10$^4$ K.  This corresponds to the net smoothing
and heating of baryons in primordial halos due to pair-instability supernovae. The mass
averaged metallicity in this case is slightly greater, $<Z> = 4.11 \times
10^{-4}$ Z$_{\odot}$.  This is due to a small net loss in baryon mass when the densities are 
smoothed.  As in Case A, peculiar velocities and dark matter density are unmodified.

The simulations are then evolved to z=3, following the evolution of the gas
using non-equilibrium chemistry and cooling from hydrogen and helium and their
ions \citep{Abel97, Anninos97}.  Metal line cooling is ignored.
At z = 7 we initialize a uniform metagalactic
ultraviolet background field \citep{Haardt96} with a slope of $q_{\alpha} = -1.8$ 
that is ramped to a pre-computed value by a redshift of z = 6. 
\citet{Zhang97} determined that such a
prescription can reproduce the Gunn-Peterson effect in the hydrogen Ly-$\alpha$
lines by a redshift of 5.5.  Self-shielding of the gas is not explicitly included
in the assumed UV background.

It is important to note that the box size, at 0.7 h$^{-1}$ Mpc, is somewhat small.  
Statistical results are reliable at $z = 15$.  However, by $z = 3$ (when the simulations are
terminated) the box is too small to be statistically reliable.  
We also performed a simulation in eight times the volume at the same mass and spatial
resolution (i.e. $512^3$ cells/particles), which gave results indistinguishable 
from what follows.
Nonetheless, all results at late times should be considered qualitative.

\section{Results}\label{results}

\begin{figure*}
\epsscale{1.1}
\plotone{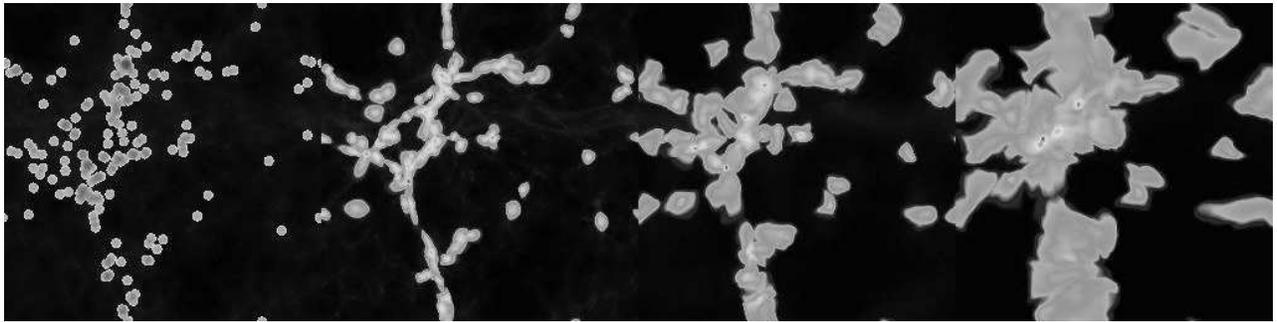}
\caption{Projected log metal density (Case A).  The area viewed is a
projection of the entire simulation volume.  The four panels
 correspond to (from left to right) z=15, 7, 5 and 3.  The Pop III supernova 
remnants are placed in the volume at z=15 and advect along the filaments.  
Photoevaporation of gas in the filaments, driven by the metagalactic UV background,
causes the volume filling factor of the metal-enriched gas to increase substantially
by z=3.}
\label{fig.pic}
\end{figure*}

\begin{figure*}
\epsscale{0.8}
\plotone{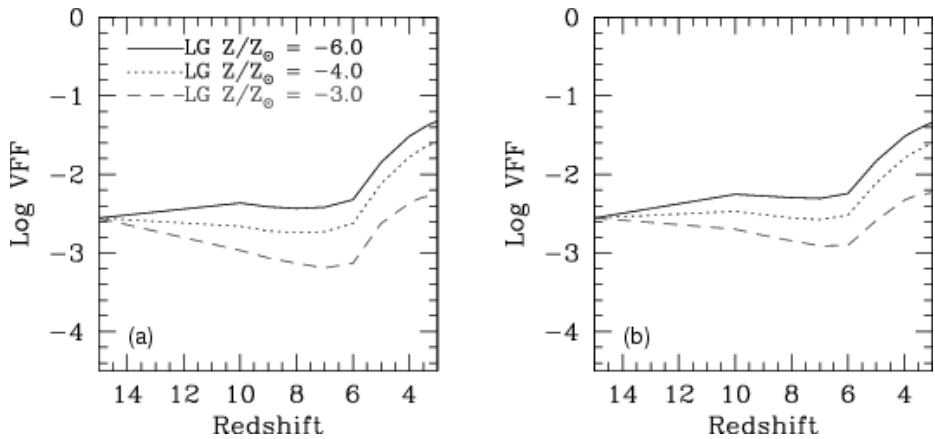}
\caption{Volume filling factor. The lines describe the
fraction of the simulation volume filled to a metallicity
of at least 10$^{-6}$ Z$_{\odot}$ (black),  10$^{-4}$ Z$_{\odot}$
(blue) and 10$^{-3}$ Z$_{\odot}$ (red).  Panel (a) corresponds to
the simulation where spheres of uniform metal density are added
and no other changes are made.
Panel (b) corresponds to the simulation where, in addition to
uniform spheres of metal density, the baryon density in the
corresponding volume is smoothed to the mean density of the
simulation and the gas temperature of the sphere is raised
to 10$^4$ K.}
\label{fig.vff}
\end{figure*}

The general evolution of our simulation after injection of metal bubbles
at $z = 15$ is as follows:  Beginning at $z = 15$, the bubbles of metal track
the flow of gas onto and along dark matter filaments.  The competing effects of
advection along the filaments and the collapse of filaments during this period
essentially cancel out, with little net effect on the fraction of the volume 
occupied by metal enriched gas (also referred to as the volume 
filling factor, or VFF).  Regions of relatively high metallicity 
($Z \geq 10^{-3} Z_{\odot}$), corresponding to the densest regions of filaments,
 decrease their volume filling factor significantly from z=15 to z=6.  Case A
shows more of a decrease in VFF at relatively high metallicities (from a VFF of
 $10^{-2.5}$ to $10^{-3.2}$) than Case B does (which has a minimum VFF of 10$^{-2.9}$)
due to the higher initial densities and lower initial temperatures of polluted 
regions in Case A.  Figure~\ref{fig.pic} shows snapshots of the metal distribution
in the simulation volume taken at 4 different redshifts, and Figure~\ref{fig.vff} 
shows the volume filling factor for three metallicity thresholds
as a function of redshift.

A uniform metagalactic ultraviolet background is switched on at $z = 7$.  
Photoheating raises the mean temperature of the baryon gas.  In the Lyman-$\alpha$
forest overdensity regime ($1 \le \delta \le 10$), which roughly corresponds
to the filamentary structure observed in Figure~\ref{fig.pic}, the temperature
is raised to $\sim$ 10,000 K.  The local thermal speed of the baryon gas then
exceeds the escape velocity of the filaments, resulting in significant 
expansion of the volume occupied by the gas in those filaments.  This includes the
gas polluted by metals.  This effect can be clearly seen in the third and
fourth panels of Figure~\ref{fig.pic}.  Figure~\ref{fig.vff} shows the
sharp increase in metal VFF in a more quantitative way.  Gas which has been
polluted above a metallicity of $10^{-6} Z_{\odot}$ (corresponding to essentially
the entire volume of gas polluted by metals) increases in VFF to 0.048 for both Case A
and 0.046 for Case B.  This corresponds to 28\% (26\%) of the baryon mass being 
enriched for Case A (Case B).  The two cases are essentially indistinguishable 
with regards to the total mass and volume of gas polluted by metals.  Examination of
gas with higher metallicity ($Z \geq 10^{-3} Z_{\odot}$) shows some difference between
the cases, with a maximum VFF of $5.2 \times 10^{-3}$ for Case A and 
$5.8 \times 10^{-3}$ for
Case B (corresponding to 1.5\% and 1.9\% of the total gas mass, respectively).  
This difference is due to the initial smoothing and heating of baryons in
Case B.

\begin{figure*}
\epsscale{0.8}
\plotone{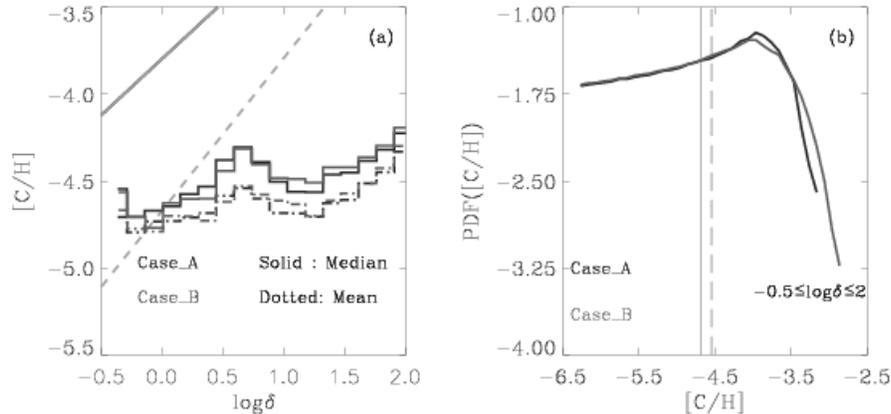}
\caption{In panel (a) we plot the volume
averaged $[C/H]$ in solar units
within constant logarithmic overdensity
bins in the range of $1 \le \delta \le 10^{2}$ at $z = 3$.
The profiles for the two cases discussed in the text show no statistical difference.
In panel (b) we plot the probability distibution function (PDF) of
$[C/H]$ within the overdensity range $1 \le \delta \le 10^{2}$.
The y-axis measures the fraction of the metal polluted volume with [C/H]
values between [C/H] and [C/H]+$\Delta$[C/H]. The result is computed with
$X_{C}=0.027$. The two distributions yield $<[C/H]> = -4.68 \pm 0.80$ for
Case A and $<[C/H]> = -4.68 \pm 0.81$ for Case B. Median values
for the two PDFs are -4.53 and -4.56 respectively.
The two cases are indistinguishable within one standard deviation.}
\label{fig.C}
\end{figure*}

In Figure~\ref{fig.C} we estimate the amount of carbon contained within the
primordial metallicity field at $z = 3$.  We assume that the carbon abundance in
the metal density is equal to $X_{C}=0.027$, which is 
taken from the supernovae 
metal yield of a massive primordial star of approximately 260 $M_{\odot}$ as
 computed by \citet{Heger02}.
  We then compute the mean and 
median carbon metallity, $[C/H] = $ log~$(n_{C}/n_{H})-[C/H]_{\odot}$
(where $[C/H]_{\odot}=-3.45$) in bins of constant logarithmic
overdensity between log~$\delta = -0.5 - 2.0$ and plot this in 
Panel A of Figure~\ref{fig.C}.  Altering the effects of $X_{C}$ results
in this figure being scaled along the y-axis by a factor of 
log$(X_{C}/0.027)$.  The solid green line is the fit to the observations 
of \citet{Schaye03} using their fiducial model. 
The dashed-green line is the fit of the lower bound lognormal scatter 
of their data.  The results of our simulation yield that the Population
III carbon content in the IGM at $z = 3$ is below the observed limits 
across the entire overdensity range.

In Panel B of Figure~\ref{fig.C} we plot the probability 
distribution function (PDF) for both simulation
cases in the overdensity range log~$\delta = -0.5$ to $+2.0$.  The vertical lines
correspond to the mean (solid) and median (dashed) values of the [C/H] values.
The distributions corresponding to the two cases are statistically 
indistinguishable within one standard deviation. 
The small variation between the two
cases in the range [C/H]=-4 to -2 is due to the difference in their
mean initial metallicity per bubble (caused by the difference
in treatment of of the baryon
density field in the region initiallly polluted by 
metals in the two cases).

In order to determine the observability of the primordial metallicity
field, we postprocess our data to compute the fraction of CIV to neutral
hydrogen (HI) for each cell in our computational volume.  Obtaining the
quantity $log~F=log~(n_{CIV}/n_{HI})$ within constant overdensity bins, in
the range $log~\delta=-0.5 - 2.0 $, allows the determination of the
lognormal average $<log~F>$ at each redshift. At z=3 our analysis yields
$<log~F> = -3.78 \pm 0.91 (-3.78 \pm 0.92)$ for Case A~(B). We can then
approximate the CIV optical depth due to PopIII stars as
$\tau_{CIV}^{popIII} \approx
(\frac{f_{CIV}\lambda_{CIV}}{f_{HI}\lambda_{HI}}) \cdot 10^{<log~F>}
\tau_{HI}$, where $f_{CIV,HI}$ and $\lambda_{CIV,HI}$ are the oscillator
strengths and rest-frame absorption wavelengths for the two species (CIV 
\& HI). Using our mean values for $<log~F>$ at z=3 we obtain an estimated
optical depth, due to CIV, of $\tau_{CIV}^{popIII} = 0.97_{0.12}^{7.86}
\times 10^{-4} ~\tau_{HI}$ ($0.97_{0.11}^{8.13} \times 10^{-4} \tau_{HI}$).  
\citet{Schaye03} computed $\tau_{CIV} = 10^{-2}~\tau_{HI}$ in the
Ly-$\alpha$ forest overdensity range. If the above value corresponds to
the total metallicity at z=3 then the contribution of the primordial
component to the total optical depth of CIV is about $\tau_{CIV}^{popIII}
= 0.01~\tau_{CIV}$.  This result is somewhat sensitive to the shape of the
UV background -- see \citet{Schaye03} for more details.
Statistical correlations between PopIII CIV and HI
absorbers and more detailed examination of spectra due to ejecta from
Population III stars will be discussed in a forthcoming paper.

\section{Discussion and Conclusions}\label{discussion}

In this paper we use cosmological hydrodynamic simulations to examine the evolution 
of metals ejected by an early population of massive primordial stars.  We
show that, in the absence of further star formation, photoevaporation of
baryons bound to dark matter filaments during reionization is the most important
mechanism in determining the volume filling fraction by $z = 3$.  Our two
study cases, although different in their initial setup, give the same results 
for the global distribution of the primordial metal field by $z = 3$, suggesting
that our result is insensitive to small-scale dynamics.

Comparison of our results to observations of carbon in the Lyman-$\alpha$ 
forest by \citet{Schaye03} show that at $z = 3$ the median value of the Pop III
carbon metallicity for both cases considered fall within the low end of the scatter 
range of the observed data for log~$\delta\le 0$.  For log~$\delta \ge 0$ the 
Population III carbon metallicity is below the observed values, with the Schaye
result showing a much stronger increase in metallicity with overdensity, resulting
in the median value of $[C/H]^{PopIII}$ becoming an increasingly smaller fraction of
the total observed [C/H].

Our results depend strongly on two factors, namely, the total number of Population III 
stars formed in our volume and the metal yield per star.  In these simulations we make the 
assumption that all halos with mass $M_{DM} \geq 5 \times 10^5 M_{\odot}$ 
form a massive primordial star by $z = 15$, which was guided by the simulations performed
by \citet{Abel02} and \citet{Yoshida03}, 
which show that this is the characteristic dark matter mass of a halo
which forms a star in the early universe.  Simulations by \citet{Machacek01} and 
semianalytical calculations by 
\citet{Wise03} show that a soft UV background produced by the first Pop III stars
effectively dissociates $H_2$, which is the primary cooling mechanism in primordial
star formation.  This so-called negative feedback
 effect raises the minimum halo mass that can form a primordial
star within it and therefore reduces the number of halos which will form 
Population III stars at a given epoch. \citet{Wise03} find that negative feedback
reduces the number of star forming halos by a factor of 5-10 relative to what we used.
On the other hand, suppression of Pop III star formation by negative feedback would be
compensated by an extended epoch of Pop III star formation.  At present, we do not know
when Pop III star formation ceases.  We view our choice of
$M_{min} = 5 \times 10^5 M_{\odot}$ at $z = 15$ as a hedge
between competing effects.

The decision to place spheres of metal in the simulation volume at $z = 15$ was guided 
primarily by the WMAP polarization results.  \citep{Kogut03}.  This choice may have
resulted in an underestimation of the number of Population III stars (and therefore
metal pollution due to Pop III supernovae) because 
there are dark matter halos which form after $z = 15$ but may still be unpolluted by metals.
However, results by \citet{bromm01} suggest the existence of a ``critical metallicity'' of
$\sim 5 \times 10^{-3} Z_{\odot}$ above which a solar IMF dominates, and it has been argued
that this metallicity is reached by $z \sim 15-20$.  \citep{Mackey03,Schneider02}  The 
choice of $z = 15$ for our epoch of instantaneous metal enrichment seems to be a reasonable 
compromise.

The physical properties of the metal ``bubbles'' can have a possible effect on our results.  The choice 
of a 1 kpc (proper) radius for the metal bubbles 
is somewhat arbitrary.  Several calculations have been performed that suggest that
ejecta from the most massive pair-instability supernovae can propagate to large distances \citep{Bromm03,Madau01},
but the maximum propagation distance is unclear.  Additionally, \citet{Bromm03} suggest that the ejecta from 
pair-instability supernovae still has substantial peculiar velocities ($\sim 50$ km/s) at 500 pc.  The metal spheres
in this calculation have no initial outward peculiar velocity, which may result in a smaller volume filling factor
than if this were taken into account.

The second factor that strongly affects our result is the choice of the amount of metals
created per Population III supernova.  \citet{Abel02} and \citet{bromm02} both suggest that
the first population of stars will be very massive.  The mass function of the first 
generation of stars is unclear,
due to lack of resolution and appropriate physics.  The main-sequence mass of the star 
strongly affects its ultimate fate:  Stars with the range of $\sim 140-260 M_{\odot}$ detonate
in pair instability supernovae, which are much more energetic (up to $\sim 10^{53}$ ergs compared
to 10$^{51}$ ergs for a standard Type I or Type II supernova) and produce more metal
 (up to 57 M$_{\odot}$ of $^{56}$Ni and almost 130 M$_{\odot}$ of total metals 
for a 260 M$_{\odot}$ primordial star).
However, stars between $\sim 50 - 140 M_{\odot}$ and above $\sim 260 M_{\odot}$ form black holes
without first ejecting significant quantities of nucleosynthesized material.  
\citep{Heger03a}.  The
amount of metals placed into the simulation volume is scalable - if the mean amount of metals 
ejected by Population III stars were lower (due to some substantial fraction collapsing 
directly into black holes, for instance), all of the results shown
in Figure~\ref{fig.vff} and Panel A of Figure~\ref{fig.C} scale linearly with the 
mean amount of metal produced per star.

Our results for [C/H] vs. overdensity (using carbon as a proxy for metallicity)
agree with the results of \citet{Schaye03} to within one standard deviation for
the lowest observed densities (log~$\delta < 0$).  These are the densities that are
the most likely to remain unpolluted by later generations of stars, which form
in deeper potential wells.  Further study of the lower density regions of the
Lyman-$\alpha$ forest could yield more constraints on the mass and total
number density of massive Population III stars.

An additional factor to consider is that the nucleosynthetic yields of very massive
primordial stars are much different than that of metal-enriched stars.  \citep{Heger03a}.
Due to this, it may be possible to disentangle the effects of massive primordial stars
and their metal-polluted descendants, as discussed by \citet{Oh01}.

Due to our choices of low minimum halo mass and high metal yield per supernova, 
our result is a strong upper limit on the pollution of the Lyman-$\alpha$ 
forest due to Population III stars, unless we have severely underestimated
the duration of the Population III epoch.

The simulation volume is relatively small.  Though a reasonable statistical 
representation of the universe at $z = 15$, the results obtained at later 
times ($z \sim 3$) should be considered qualitative due to the small box size.
A much larger simulation volume is required for adequate
statistics at $z \sim 3$.  However, simulating a much larger volume which would 
still have reasonable spatial and dark matter mass resolution on a single grid is 
computationally prohibitive at the present time.

All of the simulation results described in this paper are performed without 
further star formation or feedback. 
Having a single episode of star formation at $z=15$ means that metal evolution
after that time is passive, whereas in reality there would be continuous star formation and
feedback. A logical extension of this work is the 
inclusion of later epochs of star formation and their resulting feedback of
metals and energy into the IGM.  These results will be presented in a 
forthcoming paper.

\acknowledgements

We would like to thank Robert Harkness for invaluable suggestions
and technical assistance, and the referee, Volker Bromm, 
for suggestions which have significantly improved the quality of this manuscript.
BWO and MLN would like to thank Tom Abel, Andrea Ferrara, Wal Sargent and 
Gerry Wasserburg for stimulating discussions.  
This work was supported in part by NASA grant NAG5-12140 and NSF
grant AST-0307690.  Computational resources
for all runs were provided by the San Diego Supercomputing Center under
grant MCA098020S.


\end{document}